\documentclass[conference]{IEEEtran}

\usepackage{graphicx}
\usepackage[space]{grffile}
\usepackage{latexsym}
\usepackage{textcomp}
\usepackage{longtable}
\usepackage{booktabs}
\usepackage{amsfonts,amsmath,amssymb}
\usepackage{url}
\usepackage[utf8]{inputenc}
\usepackage[ngerman,english]{babel}

\usepackage{amssymb}

\usepackage{nicefrac}

\usepackage{amsmath}
\usepackage{amsfonts}

\newcommand{\calG}{\mathcal{G}}
\newcommand{\calN}{\mathcal{N}}
\newcommand{\calE}{\mathcal{E}}

\begin{document}

\title{Uncovering the Spread of Chagas Disease in Argentina and Mexico}

\author{
\IEEEauthorblockN{
Juan de Monasterio\IEEEauthorrefmark{1},
Alejo Salles\IEEEauthorrefmark{1}\IEEEauthorrefmark{2},
Carolina Lang\IEEEauthorrefmark{1},
Diego Weinberg\IEEEauthorrefmark{3}, 
Martin Minnoni\IEEEauthorrefmark{4}, 
Matias Travizano\IEEEauthorrefmark{4},  \\
Carlos Sarraute\IEEEauthorrefmark{4}
}
\IEEEauthorblockA{\IEEEauthorrefmark{1}FCEyN, Universidad de Buenos Aires, Argentina. Email:\emph{ \{laterio, carolinalang93\}@gmail.com}}
\IEEEauthorblockA{\IEEEauthorrefmark{2}Instituto de C\'alculo and CONICET, Argentina. Email: \emph{alejo@df.uba.ar} }
\IEEEauthorblockA{\IEEEauthorrefmark{3}Fundación Mundo Sano, Buenos Aires, Argentina. Email: \emph{dweinberg@mundosano.org} } 
\IEEEauthorblockA{\IEEEauthorrefmark{4}Grandata Labs, 550 15th Street, San Francisco, CA, USA. \\ Email:\emph{ \{martin, mat, charles\}@grandata.com }
} 
}

\maketitle

\begin{abstract}
Chagas disease is a neglected disease, 
and information about its geographical spread is very scarse.
We analyze here mobility and calling patterns in order to identify potential risk zones for the disease,
by using public health information and mobile phone records. 
Geolocalized call records are rich in social and mobility information, which can be used to infer whether an individual has lived in an endemic area. 
We present two case studies in Latin American countries.
Our objective is to generate risk maps which can be used by public health campaign managers to prioritize detection campaigns and target specific areas. 
Finally, we analyze the value of mobile phone data to infer long-term migrations, which play a crucial role in the geographical spread of Chagas disease.
\end{abstract}

\section{Introduction}

Chagas disease is a neglected tropical disease of global reach, spread mostly across 21 Latin American countries. 
Caused by the \textit{Trypanosoma cruzi} parasite, its transmission occurs mostly in the American endemic regions via the \textit{Triatoma infestans} insect family. 
In recent years, due to the globalization of migrations, the disease has become an
issue in other continents as well~\cite{schmunis2010chagas}, 
particularly in countries that receive Latin American immigrants 
such as Spain and the United States.

A crucial characteristic of the infection is that it may last 10 to 30 years in an individual without presenting symptoms~\cite{rassi2012american}, which greatly complicates effective detection and treatment. 
Long-term human mobility (particularly seasonal and permanent rural-urban migration) thus plays a key role in the geographical spread of the disease~\cite{briceno2009chagas}.

In this work, we discuss the use of Call Detail Records (CDRs) for the analysis of mobility patterns and the detection of potential risk zones of Chagas disease in two Latin American countries~\cite{deMonasterio2016analyzing,Sarraute2015descubriendo}.
We generate predictions of population movements between different regions, providing a proxy for the epidemic spread. 
We present two case studies, in Argentina and in Mexico, using data provided by mobile phone companies from each country.

\section{Chagas Disease in Argentina and Mexico}

\subsection{Endemic Zone in Argentina}\label{endemic_zone_argentina}

The \textit{Gran Chaco}, situated in the northern part of the country,
is endemic for the disease~\cite{OPS2014mapa}. 
The ecoregion's low socio-demographic conditions further support the parasite's lifecycle, where domestic interactions between humans, triatomines and animals foster the appearance of new infection cases, particularly among rural and poor areas.
This region is considered as the endemic zone $E_Z$ in the analysis described in Section~\ref{riskmaps}.

Recent national estimates indicate that between 1.5 and 2 million individuals carry the parasite, with more than seven million exposed. 
National health systems face many difficulties to effectively treat the disease. 
In Argentina, less than 1\% of infected people are treated 
(the same statistic holds at the world level).
Even though governmental programs have been ongoing for years now,
data on the issue is scarce or hardly accessible.

\subsection{Endemic Zone in Mexico} \label{endemic_zone_mexico}

The Mexican epidemic area~\cite{cruz2006chagmex}
covers most of the South region of the country and includes the states of Jalisco, Oaxaca, Veracruz, Guerrero, Morelos, Puebla, Hidalgo and Tabasco.
This region is considered as the endemic zone $E_Z$ for the Mexican case.

Despite the lack of official reports, an estimate of the number of \textit{Trypanosoma cruzi} infections by state in the country
indicates that the number of potentially
affected people in Mexico is about 5.5 million~\cite{carabarin2013chagas}.
In recent years there has been a focus on treating the disease with two available
medications, benznidazole or nifurtimox, 
with less than 0.5\% of infected individuals receiving treatment in Mexico~\cite{manne2013barriers}.

People from endemic areas tend to migrate to the industrialized cities of the country, mainly Mexico City, in search for jobs~\cite{guzman2001epidemiology}. 
Therefore, the study of long-term mobility is crucial to understand the geographical spread of the Chagas disease in Mexico.

\section{Mobile Phone Data Sources}

Our data source is anonymized traffic information from two mobile operators.
The Argentinian dataset contains CDRs collected over a period of 5 consecutive months.
The Mexican dataset contains CDRs for a period of 24 consecutive months.

For our purposes, each record is represented as a tuple $\left < i, j, t, d, l \right >$,
where user $i$ is the encrypted caller, user $j$ is the encrypted callee, $t$ is the date and time of the call,
$d$ is the direction of the call (incoming or outgoing), and $l$ is the location of the tower that routed the communication.
The dataset does not include personal information from the users.

We aggregate the call records for a five 
month period into an edge list $(n_i, n_j, w_{i,j})$ where nodes $n_i$ and $n_j$ 
represent users $i$ and $j$ respectively, and $w_{i,j}$ is a boolean value
indicating whether those two users have communicated during the 
five month period. 
This edge list represents our communication graph  
$\calG = \left< \calN, \calE \right> $ where $\calN$ denotes the set of nodes (users) 
and $\calE$ the set of communication links. 
We note that only a subset $\calN_C$ of nodes in $\calN$
are clients of the mobile operator.
Since geolocation information is available only for users in $\calN_C$, in the analysis we considered the graph $\calG_C = \left< \calN_C, \calE_C \right> $ of communications between clients of the operator.

\section{Risk Maps for Chagas Disease}\label{riskmaps}

\subsection{Methodology for Risk Map Generation} \label{methods}

The first step is to determine the area where each user lives. 
For each user $u \in \calN_C$, we compute its \textit{home antenna} $H_u$ as the antenna in which user $u$ spends most of the time during weekday nights~\cite{csaji2012exploring}.
The users such that $H_u$ is located in the endemic zone $E_Z$ are considered the \textit{residents of $E_Z$}.

The second step is to find users highly connected with the residents of $E_Z$. 
To do this, we compute the list of calls for each user and then determine his set of neighbors in the social graph $\calG_C$. For each resident of the endemic zone, we tag all his neighbors as \textit{vulnerable}.
    
    The third step is to aggregate this data by antenna. For every antenna $a$, we compute:
        the total number of residents $N_a$,
        the total number of residents which are vulnerable $V_a$,
	the total volume of outgoing calls $C_a$,
	and the number of outgoing calls whose receiver lives in the endemic area $VC_a$. 

    We generated heatmaps to visualize these antenna indicators, overlapping these heatmaps with political maps. 
Each antenna is represented by a circle whose
		\textit{area} depends on the population living in the antenna $N_a$
		and whose \textit{color} depends of the fraction ${V_a}/{N_a}$ of vulnerable users living there.
    We used two filtering parameters:
        each antenna is plotted if its fraction of vulnerable users is higher than $\beta$,
        and if its population is bigger than $m_v$.

\subsection{Results and Observations} \label{results}

\begin{figure}[t]

\begin{minipage}{.495\linewidth}
\centering
  \includegraphics[width=0.90\linewidth]
  {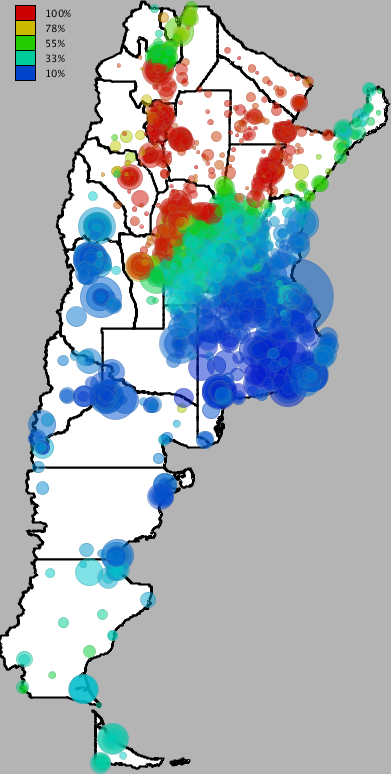}
  
(a) $\beta = 0.01$
\end{minipage}
\begin{minipage}{.495\linewidth}
\centering
  \includegraphics[width=0.90\linewidth]
  {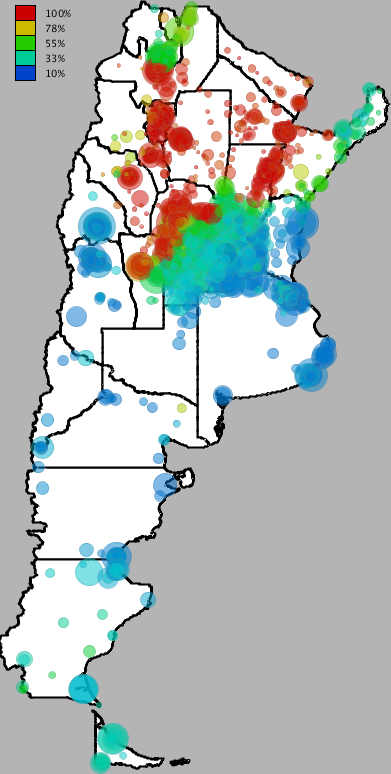}
  
(b) $\beta = 0.15$
\end{minipage}
\caption{Risk map for Argentina, filtered according to $\beta$.}
\label{fig:mapa_argentina}
\end{figure}

Fig.~\ref{fig:mapa_argentina} shows the risk maps for Argentina, generated with
two values for the $\beta$ parameter and fixing $m_v = 50$ inhabitants per antenna. After filtering with $\beta = 0.15$, we see that large portions of the country harbor potentially vulnerable individuals.
Namely, Fig.~\ref{fig:mapa_argentina}(b) shows antennas where more that 15\% of the population has social ties with the endemic region $E_Z$.

\begin{figure}[b]
\centering
\includegraphics[width=0.70\linewidth]
{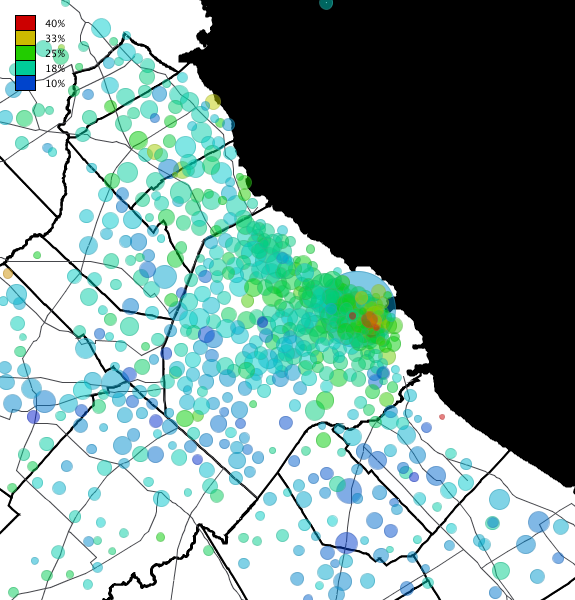}
\caption{Risk map for the metropolitan area of Buenos Aires, filtered with $\beta = 0.02$.}
\label{fig:amba_map}
\end{figure}

Advised by Mundo Sano Foundation's experts, we then focused on areas whose results were unexpected to the epidemiological experts. 
Focused areas included the provinces of Tierra del Fuego, Chubut, Santa Cruz and Buenos Aires, with special focus on the metropolitan area of Greater Buenos Aires whose heatmap is shown in Fig.~\ref{fig:amba_map}.

High risk antennas were separately listed and manually located in political maps. This information was made available to the Mundo Sano Foundation collaborators who used it as an aid for their campaign planning and for the education of community health workers.

\section{Prediction of Long-term Migrations} \label{long_term}

In this section, we describe our work on the prediction of long-term mobility.
The CDR logs available in the Mexican dataset span 24 months, from January 2014 to December 2015, making them suitable for this study.

We divide the available data into two distinct periods:
$T_0$, from January 2014 to July 2015, considered as the ``past" in our experiment;
and $T_1$, from August 2015 to December 2015, considered as the ``present".
Knowing which users live in the endemic region $E_Z$ and how they communicate
during period $T_1$,
we want to infer whether they lived in $E_Z$ in the past (period $T_0$).

\subsection{Model Features}

The features constructed reflect calling and mobility patterns.
Each week is divided into 3 time periods: (i) \textit{weekday} from Monday to Friday, on working hours (8hs to 20hs); (ii) \textit{weeknight} from Monday to Friday, between 20hs and 8hs of the following day;
and (iii) \textit{weekend} is Saturday and Sunday.
The model consists of the following features, which can be classified in 4 categories:

\subsubsection{Used and home antennas}\label{homeantenna}
For each user $u \in \calN_C$, we register the top ten most used antennas, 
considering all calls or only calls made during the \textit{weeknight} period.
Users were tagged as `endemic' if their home antenna is in the endemic zone $E_Z$ and `exposed' if any of the top ten antennas is in the risk area.

\subsubsection{Mobility diameter}
The user's logged antennas define a convex hull in space and the radius of the hull is taken as the mobility diameter. 
We generate two values, considering (i) all antennas and (ii) only the antennas used during the \textit{weeknight}.

\subsubsection{Communications graph}

We enrich the social graph $\calG_C$ built from the CDRs.
For each edge $\left< n_i, n_j \right> \in \calE_C$, 
we gather the number of calls exchanged, the sum of call durations (in seconds), the direction (incoming or outgoing),
segmented according to the periods \textit{weekday}, \textit{weeknight}, and \textit{weekend}.

Since the samples in our dataset are users, we aggregate these variables by grouping interactions at the user level. The combination of different variables amounts to a total of 130 features per user.
 
We also compute the user's degree and the total count of endemic neighbors, labeling each user $i$ as \textit{vulnerable} whenever he has an edge with a user $j$ who lives in the endemic region $E_Z$.

\subsubsection{Validation data} 

We perform an analysis similar to the home antenna detection previously described, 
but considering the time period $T_0$ (from January 2014 to July 2015),
in order to determine the home antenna of users during $T_0$.

\subsection{Supervised Classification}

We used most common techniques for this task:
Support Vector Machines, Random Forest, Logistic Regression, and Multinomial Naive Bayes.
The data was split into 70\% for training and 30\% for testing.

The Multinomial Bayes classifier has a linear time complexity, and thus serves as a fast benchmark.
Support Vector Machines (SVM) and Logistic Regression
performed better than Multinomial Bayes. 
We tuned the standard hyperparameters: $L2$-penalty regularization for Logistic Regression and kernel bandwidth for the Gaussian Kernel SVM.
Both learning routines were executed in parallel and in each iteration 5\% of the training set was sampled for cross validation.
The best model was a Logistic Regression Classifier with an $L2$-penalty value of 0.01. 
The scores obtained by the selected model on the out-of-sample set are
F1-score: 0.964537; 
accuracy:  0.980670;
AUC: 0.991593;
precision: 0.970838;
recall: 0.958316.

High values across all scoring measures are achieved. 
These results can be explained by the fact that 
communication and mobility patterns are in essence highly correlated across time periods. 
In this case,
a user being endemic in $T_1$ is correlated to being endemic in $T_0$, and the same holds with a user's interaction with vulnerable neighbors during $T_1$.

\section{Conclusion}

The heatmaps shown in Section~\ref{riskmaps} expose an expected ``temperature'' descent from the endemic regions outwards. 
We also found out communities atypical compared to their neighboring region, which stand out for their strong communication ties with the endemic region $E_Z$.
The detection of these communities is of great value to health campaign managers, providing them tools to target specific areas and prioritize resources and calls to action more effectively.

In Section~\ref{long_term}, we tackled the problem of predicting long-term migrations. In particular, we showed that it is possible to use the mobile phone records of users during a bounded period of time in order to predict whether they have lived in the endemic zone $E_Z$ in a previous time frame.

To conclude, we showed here the value of generating risk maps in order to prioritize effectively detection and treatment campaigns
for the Chagas disease. 
The results stand as a proof of concept which can be extended to other countries with similar characteristics.

\bibliographystyle{abbrv}
\bibliography{../bibliography/epidemics}

\end{document}